\documentclass{emulateapj}

\usepackage[latin1]{inputenc}
\usepackage[english]{babel}

\newcommand{\ptsla}{Philos. Trans. R. Soc. London A.}

\newcommand{\tap}{IEEE Trans. Antennas Propagation}
\renewcommand{\jqsrt}{J. Quant. Spec. Radiat. Transf.}

\begin{document}

\title{High-$J$ $v$=0 SiS Maser Emission in IRC+10216: A New Case of Infrared Overlaps}

\author{J. P. Fonfr\'{\i}a Exp\'osito, M. Ag\'undez, B. Tercero,
J. R. Pardo and J. Cernicharo}

\affil{Dept. Molecular and Infrared Astrophysics, Instituto
de Estructura de la Materia, CSIC, C/ Serrano 121, 28006, Madrid
(Spain)}

\begin{abstract}

We report on the first detection of maser emission in the $J$=11-10,
$J$=14-13 and $J$=15-14 transitions of the $v$=0 vibrational state
of SiS toward the C-rich star IRC+10216. These masers seem to be
produced in the very inhomogeneous region between the star and
the inner dust formation zone, placed at $\simeq$5-7~R$_*$, with expansion
velocities below 10~km~s$^{-1}$. We interpret the pumping mechanism
as due to overlaps between $v$=1-0 ro-vibrational lines of
SiS and mid-IR lines of C$_2$H$_2$, HCN and their $^{13}$C
isotopologues. The large number of overlaps found suggests the
existence of strong masers for high-$J$ $v$=0 and $v$=1 SiS
transitions, located in the submillimeter range. In addition, it could be
possible to find several rotational lines of the SiS isotopologues
displaying maser emission.

\end{abstract}


\section{Introduction}

The detection of strong maser emission at the frequencies of pure
rotational transitions of some molecules is a common phenomenon in
circumstellar envelopes (CSE's) of evolved stars
\citep{elitzur_1992,gray_1999}. The maser is usually produced in a
small region of the envelope and sometimes provides valuable
information on the physical conditions of the emitting region.

Due to the different chemistry, masers are
produced by different molecules in O- and C-rich
stars.
In O-rich stars, SiO exhibits strong maser emission in
different rotational transitions within several vibrational
states, from $v$=1 to 4 \citep{cernicharo_1993,pardo_1998}.
These masers
are formed in a region of the CSE very close to the stellar
surface and seem to be driven by NIR radiation
\citep{pardo_2004}.
In C-rich stars,
although SiO is present with similar abundances than in O-rich
stars \citep{schoier_2006}, no SiO maser has been detected.
The explanation could be that SiO is formed at
$\simeq$3-5~R$_*$,
where the angular dilution of the star is high and the density
and temperature lower than in the regions where SiO masers
are produced in O-rich stars
\citep{agundez_2006}. In C-rich stars
only HCN shows strong maser emission in several pure rotational
lines within vibrational states from $\nu_2$=1 to 4
\citep{lucas_1989,schilke_2003}.
These masers must be formed in the innermost regions of
the CSE.
SiS has been previously found to show weak maser
emission in the $J$=1-0 $v$=0 transition in IRC+10216
\citep{henkel_1983}.

In this letter, we report on the first detection of maser
emission from the $J$=11-10, 14-13 and 15-14 transitions
in the $v$=0 vibrational state of SiS (hereafter M$_1$, M$_2$, and
M$_3$) observed toward the C-rich star IRC+10216. We have also
obtained
observations of $v$=1 rotational lines which exhibit
thermal emission. We propose that overlaps of $v$=1-0
ro-vibrational transitions of SiS with mid-IR lines of
C$_2$H$_2$ and HCN could provide the pumping
mechanism for these masers as well as higher-$J$ $v$=0 SiS
masers in the submillimeter range. This discovery
is interesting because this species could play in C-rich
stars a role similar to that of SiO in O-rich stars:
the energy level
pattern of both molecules is similar and
it is also formed close to the star, as chemical equilibrium
and interferometric observations imply \citep{bieging_1989}.

\section{Observations}

\begin{figure*}
\epsscale{0.865}
\plotone{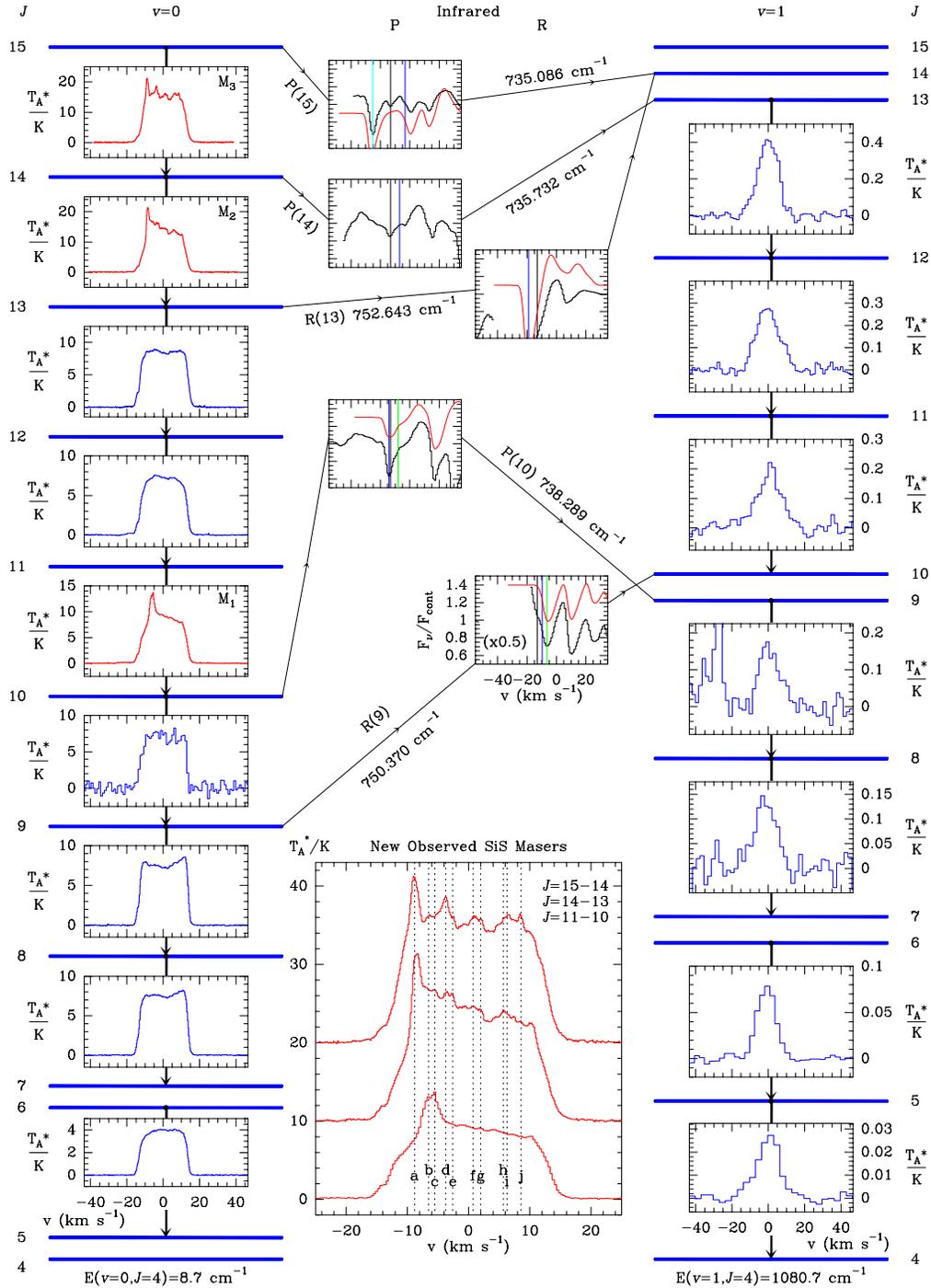}
\caption{Millimeter and mid-IR observations towards
IRC+10216. -- The pure rotational transitions in the $v$=0 state
(left inserts) and in the $v$=1 state (right inserts) are plotted
as a function of the velocity with respect to the star
(v$_{LSR}$=$-26.5$~km~s$^{-1}$). The velocity structure of the
three observed masers M$_1$, M$_2$ and M$_3$
(plotted as red histograms) is shown in more detail in the
middle-bottom insert, where the upper lines have been increasingly
shifted by 10~K. -- The middle inserts present some mid-IR
spectra at the frequencies of several ro-vibrational $v$=1-0 SiS
lines, whose labels and frequencies are also shown. The
zero of the velocity scale is set to the SiS frequency corrected
for the LSR. The observed spectrum is plotted as a black
histogram. Red lines correspond to a radiative transfer model
which only considers C$_2$H$_2$, HCN and their isotopologues
\citep{fonfria_2006}. The vertical lines indicate the
position of maximum absorption for the transitions of SiS
(black), C$_2$H$_2$ (dark blue), H$^{13}$CCH (light blue) and HCN
(green).} \label{fig:figure}
\end{figure*}

The observations of the $v$=0 $J$=6-5 and $J$=8-7 to $J$=15-14
transitions of
SiS (see left panels, Fig.~\ref{fig:figure}) were carried out on
2004 June 19$^\textnormal{\scriptsize{th}}$ with the IRAM 30 m radio telescope. Four
SIS receivers operating at 1, 1.3, 2, and 3~mm were used
simultaneously. System temperatures were in the range
120-225~K for the 1, 1.3
and 2~mm receivers and 200-600~K for the 1~mm receivers. Atmospheric
opacities ranged between 0.08 at 108~GHz to 0.28 at 267~GHz. For the
$J$=10-9 SiS line at 181.5~GHz, the system temperature was
significantly higher, $\simeq$10$^{4}$~K, due to proximity to
the atmospheric water
line at 183.3~GHz. The intensity scale was calibrated using two
absorbers at different temperatures according to the Atmosphere
Transmission Model ATM \citep{cernicharo_1985,pardo_2001}.
Pointing and focus were regularly checked on the nearby quasar OJ
287. The observations were made in wobbling mode, with 
180'' offset and the secondary
nutating at a rate of 0.5~Hz. The backends were a filter bank
with 256~MHz bandwidth and 1~MHz resolution and an autocorrelator
with 80~kHz resolution ($\Delta$$v$$\simeq$0.08$-$0.2~km~s$^{-1}$).
The lines of SiS in the $v$=1 state (see right panel,
Fig.~\ref{fig:figure}) were also observed with the IRAM 30~m telescope
between 1989 and 2001.

Additionally we show some spectra corresponding to ro-vibrational
lines of SiS (see center panels in Fig.~\ref{fig:figure}) from a
mid-IR high resolution spectrum (11-14~$\mu$m) obtained in 2002
December with the TEXES spectrometer mounted on the 3~m
telescope IRTF
\citep*[observational procedures described in][]{lacy_2002,fonfria_2006}.

\section{Results and Discussion}

The $v$=1 rotational lines show single cusped profiles and their
relative intensities indicate that the rotational levels are
thermally populated. The linewidths correspond to
velocities of 9-11~km~s$^{-1}$, lower than the terminal velocity
$\simeq$14.5~km~s$^{-1}$ \citep{cernicharo_2000},
thus the emission arises from the
innermost region of the CSE, between the photosphere and the inner
dust formation zone, placed at $\simeq$5~R$_*$ \citep{keady_1993}.
We can derive the SiS abundance in that region from the
$v$=1 lines assuming an uniform sphere at a distance of 180 pc
with a radius of 10 R$_*$, illuminated by the central star
(R$_*$=5$\times$10$^{13}$ cm, T$_*$=2300~K), with
T$_\textnormal{\scriptsize{k}}$=1000~K and n$_{\textnormal{\scriptsize{H}}_2}$=1.6$\times$10$^{9}$~cm$^{-3}$ (mean
values for this region derived by \citealt{fonfria_2006}) and an
expansion velocity of 11~km~s$^{-1}$.
We solve the statistical
equilibrium for SiS considering 100 rotational levels and 3
vibrational states and apply the LVG radiative transfer formalism
using the code developed by J. Cernicharo.
With the assumed n$_{\textnormal{\scriptsize{H}}_2}$, the column density for H$_2$ is
$\simeq$7$\times$10$^{23}$~cm$^{-2}$ and the derived one for
SiS is $\simeq$5.0$\times$10$^{18}$~cm$^{-2}$. Hence, the SiS abundance is
$\simeq$7$\times$10$^{-6}$.
This result is compatible with a higher SiS abundance in the innermost
CSE (from LTE chemistry models,
3$\times$10$^{-5}$, \citealt{agundez_2006})
and an abundance further away of 6.5$\times$$10^{-7}$ according to observations
of $v$=0 rotational lines over
the outer CSE by \citet{bieging_1989}.

Most $v$=0 rotational lines show a rounded or slightly double
peaked profile with the blue part absorbed by cold SiS through
the envelope.
However, M$_1$, M$_2$ and M$_3$
show extra emission in the form of narrow peaks (FWHM=1-3~km
s$^{-1}$). The velocities of these features, within the $-10$ to 10~km
s$^{-1}$ range, indicate that SiS maser emission arises from different regions
located between the star and the inner dust formation zone
($r$$\simeq$5~R$_*$). The lines M$_2$ and M$_3$ have been previously
observed by \citet{sahai_1984} but no maser emission was
noticed. This could be due either to the limited sensitivity
of their observations or to a time variability of the SiS maser
phenomenon.
The bottom-center panel of Fig.~\ref{fig:figure} shows in
detail the line profiles of the observed masers. Up to ten maser
features labelled ($a$,\ldots,$j$) are identified.
The most complex line profile is
that of M$_3$. It is formed by 5 main features: $a$, $d$, $f$,
$i$, $j$ with $v$=$-8.8$, $-3.7$, $0.88$, $6.3$ and $8.4$~km
s$^{-1}$, respectively.
The line profiles show
that the strongest peaks are at negative velocities, having their
red counterparts rather weak. This behavior was previously found
by \citet{henkel_1983} for the $v$=0 $J$=1-0 line, which mostly
consists of a narrow peak (FWHM=0.3~km~s$^{-1}$) centered at
$-13.5$~km~s$^{-1}$. The strong asymmetry of the
lineshapes can be either due to blanking by the star of the
redshifted maser or by amplification of the blueshifted emission by
the foreground stellar environment.

The strongest observed maser, M$_3$, with
T$_\textnormal{\scriptsize{MB}}$$\simeq$60~K
(F$_\textnormal{\scriptsize{obs}}$$\simeq$300~Jy; with thermal
and non-thermal emission), is weak compared to some SiO
masers detected in O-rich stars \citep*[e.g.][]{cernicharo_1993}
or to HCN masers observed in IRC+10216 \citep{lucas_1989,schilke_2003}.
However, M$_1$, M$_2$, and M$_3$ are stronger than
the SiS $J$=1-0 maser observed towards IRC+10216 by \citet{henkel_1983}.
The similarity between maser features in
M$_2$ and M$_3$ indicates that they may arise from the same
regions and produced by the same pumping mechanism; the
maser in M$_1$ is probably
formed in other regions. Hence, we suggest two possible
geometries of the innermost CSE to explain the observed features:

($i$) An onion-like innermost region, where each maser is
produced in a shell. This
hypothesis is supported by the symmetry of
features $a$--$j$ and $d$--$i$. The peaks at
extreme velocities, $a$ and $j$, would be produced just in front of
and behind the star near the inner
dust formation shell ($r$$\simeq$5~R$_*$) with expansion velocities of
$\simeq$5-11~km~s$^{-1}$.
The features
$d$ and $i$ would be produced in a similar way but in an inner
shell with a lower expansion velocity. Finally, the central peak,
$f$, would be formed in a shell very close to the star, with the
whole shell contributing to the maser emission.
M$_1$ would be produced in a cap-shaped region in front of
the star.

($ii$) All the masers are formed in different positions of
a clumpy shell. The different features in M$_2$ and M$_3$ would
be produced in different regions of the shell: peaks $a$ and
$j$ in front of and behind the star and the other peaks
($d$, $f$, and $i$) in different clumps, as occur with the only
feature of M$_1$.

The classic pumping mechanism for the SiO $v$$>$0 masers observed
in O-rich stars resides in the increase of the trapping lifetime
(A/$\tau$)$_{v \rightarrow v-1}$ with
$J$ for $v$$\rightarrow$$v$-1 transitions, when they become
optically thick \citep{kwan_1974}. Such mechanism produces masers
in adjacent rotational lines of the $v$ state, and explains the $v$=1 and
2 SiO masers \citep{bujarrabal_1981,lockett_1992}.
However, the masers observed
in rotational transitions of $^{29}$SiO, $^{30}$SiO, and in $v$=3
and 4 of SiO do not show the latter behavior and
have been interpreted as due to IR
overlaps between ro-vibrational lines of SiO isotopologues
\citep{cernicharo_1991,gonzalezalfonso_1997}.
For SiS, the absence of maser emission in $v$=1 rotational lines
and the odd $v$=0 pattern also exclude the
Kwan \& Scoville pumping mechanism. This suggests that overlaps of
$v$=1-0 ro-vibrational transitions of SiS
with those of mode $\nu_5$ of C$_2$H$_2$
and mode $\nu_2$ of HCN, could provide the
pumping mechanism. C$_2$H$_2$ and HCN
are abundant in the inner CSE of IRC+10216 and dominate the 11-14 $\mu$m
spectrum \citep{fonfria_2006}. Overlaps with these two species
have been already proposed by \citet{sahai_1984} to
explain the different profiles of
adjacent $J$ lines of SiS. However, the SiS frequencies used in that work
were not as accurate ($\sigma$$\sim$10$^{-1}$~cm$^{-1}$) as those
available today. We have calculated those frequencies from the Dunham
coefficients determined by \citet{sanz_2003}, for which the error
of the band center is $<$$10^{-4}$~cm$^{-1}$
($\simeq$0.04~km~s$^{-1}$; the relative accuracy of P and R
lines is much better).
The frequencies of C$_2$H$_2$, H$^{13}$CCH, HCN, and H$^{13}$CN
lines have been taken from the HITRAN Database 2004
\citep{rothman_2005}, with an accuracy better than $10^{-3}$
cm$^{-1}$ ($\simeq$0.4~km~s$^{-1}$) for C$_2$H$_2$ and H$^{13}$CCH,
and $10^{-4}$~cm$^{-1}$ for HCN and H$^{13}$CN.

Table~\ref{tab:frequencies_sis} shows the mid-IR line
overlaps of SiS with C$_2$H$_2$, HCN, and their most abundant
isotopologues. For the overlap search we selected coincidences
within $|\Delta v|$$<$10~km~s$^{-1}$. However, since the
CSE is expanding, every region of the envelope is receding from
the others. Hence, if the population of the SiS levels is
affected by an overlap with a strong line of other
species, the frequency of this overlapping transition must be
higher than the SiS one. This would restrict the condition to
positive $\Delta v$.
Nevertheless, due to the linewidth, lines at
$\Delta v$$<$0 can overlap the SiS lines.
Hence, we have set the negative cutoff
to one half of the typical linewidth in the
innermost CSE ($\simeq$5~km~s$^{-1}$; \citealt{fonfria_2006}).
Therefore, our search is restricted to
$-2.5\le\Delta v\le 10$~km~s$^{-1}$.
All the ro-vibrational SiS lines commented
hereafter refer to $v$=1$\rightarrow$0 transitions and will be
labelled with the usual spectroscopic nomenclature R, Q, P
(see footnote of Table~\ref{tab:frequencies_sis}).

\begin{deluxetable}{c@{}c|c@{}c@{}c}
\tabletypesize{\scriptsize} \tablewidth{0pc} \tablecolumns{5}
\tablecaption{Mid-infrared Line Overlaps of SiS with C$_2$H$_2$,
HCN, and Their Most Abundant Isotopologues}
\tablehead{\colhead{Line} & \colhead{$\nu$
(cm$^{-1}$)} & \colhead{Mol.} & \colhead{Transition} & \colhead{$\Delta
v$(km/s)}} \startdata
\multicolumn{5}{c}{\textbf{Overlaps Involving SiS Levels of Observed Lines}}\\[2pt]
R 9 & 750.3695 & HCN         &  $01^{1}0-00^{0}0$ R$_e\left(12\right)$        & $-6.6$  \\
P10 & 738.2889 & C$_2$H$_2$  &  $1^{-1}1^{1}-1^{-1}0^{0}$ R$_f\left(3\right)$ & \phs{}1.2  \\
R13 & 752.6431 & C$_2$H$_2$  &  $0^{0}1^{1}-0^{0}0^{0}$ R$_e\left(9\right)$   & \phs{}5.9  \\[2pt]
P14 & 735.7324 & C$_2$H$_2$  &  $0^{0}1^{-1}-0^{0}0^{0}$ Q$_e\left(38\right)$ & $-6.1$  \\
P15 & 735.0861 & C$_2$H$_2$  &  $0^{0}1^{-1}-0^{0}0^{0}$ Q$_e\left(36\right)$ & $-9.9$  \\[2pt]
\multicolumn{5}{c}{\textbf{Overlaps Involving SiS Levels of Unobserved Lines}}\\[2pt]
P 2 & 743.2624 & C$_2$H$_2$  &  $0^{0}1^{1}-0^{0}0^{0}$ R$_e\left(5\right)$    & \phs{}0.6 \\
P16 & 734.4368 & C$_2$H$_2$  &  $0^{0}1^{-1}-0^{0}0^{0}$ Q$_e\left(34\right)$  & \phs{}1.1 \\
P20 & 731.8114 & C$_2$H$_2$  &  $0^{0}1^{-1}-0^{0}0^{0}$ Q$_e\left(24\right)$  & \phs{}9.5 \\
P22 & 730.4815 & C$_2$H$_2$  &  $1^{-1}1^{1}-1^{1}0^{0}$ Q$_e\left(18\right)$  & $-2.1$ \\
R22 & 757.5819 & C$_2$H$_2$  &  $1^{1}1^{1}-1^{1}0^{0}$ R$_e\left(10\right)$   & \phs{}0.1 \\
R22 & 757.5819 & H$^{13}$CN  &  $01^{1}0-00^{0}0$ R$_e\left(17\right)$         & \phs{}5.2 \\
P23 & 729.8123 & H$^{13}$CCH &  $0^{0}1^{-1}-0^{0}0^{0}$ Q$_e\left(19\right)$  & \phs{}8.7 \\
P24 & 729.1402 & C$_2$H$_2$  &  $0^{0}1^{-1}-0^{0}0^{0}$ Q$_e\left(1\right)$   & \phs{}9.4 \\
P25 & 728.4654 & H$^{13}$CCH &  $0^{0}1^{-1}-0^{0}0^{0}$ Q$_e\left(7\right)$   & $-0.0$ \\
R25 & 759.1733 & HCN         &  $01^{1}0-00^{0}0$ R$_e\left(15\right)$         & \phs{}3.6 \\
R26 & 759.6977 & C$_2$H$_2$  &  $0^{0}1^{1}-0^{0}0^{0}$ R$_e\left(12\right)$   & $-1.0$ \\
R33 & 763.2816 & H$^{13}$CN  &  $01^{1}0-00^{0}0$ R$_e\left(19\right)$         & \phs{}6.1 \\
P38 & 719.4363 & C$_2$H$_2$  &  $1^{1}1^{1}-1^{1}0^{0}$ P$_e\left(5\right)$    & \phs{}4.5 \\
R40 & 766.7130 & C$_2$H$_2$  &  $0^{0}1^{1}-0^{0}0^{0}$ R$_e\left(15\right)$   & \phs{}4.3 \\
R42 & 767.6652 & C$_2$H$_2$  &  $1^{1}1^{-1}-1^{1}0^{0}$ R$_e\left(20\right)$  & \phs{}8.7 \\
R45 & 769.0697 & C$_2$H$_2$  &  $0^{0}1^{1}-0^{0}0^{0}$ R$_e\left(16\right)$   & $-1.8$ \\
P48 & 712.1720 & C$_2$H$_2$  &  $1^{-1}1^{-1}-1^{-1}0^{0}$ P$_f\left(8\right)$ & \phs{}4.7 
\enddata
\tablecomments{Overlaps of SiS with C$_2$H$_2$, H$^{13}$CCH, HCN
and H$^{13}$CN found in the mid-IR range
$\left[690,780\right]$~cm$^{-1}$ with $-2.5\le\Delta v\le 10$
km~s$^{-1}$, where $\Delta v/c=[\nu (\textnormal{X})-\nu
(\textnormal{SiS})]/ \nu (\textnormal{SiS})$ and $J\le 50$,
(corresponding SiS $v$=0 rotational
frequencies below 900~GHz). The lines and frequencies at the left correspond
to the vibrational transitions $v$=1$\rightarrow$0 of SiS. The
errors on the velocities are $< 0.5$~km~s$^{-1}$. The notation for
the C$_2$H$_2$ and H$^{13}$CCH vibrational states involved in the
ro-vibrational transitions is $\nu_4^{\ell_4}\nu_5^{\ell_5}$,
whereas for HCN and H$^{13}$CN is $\nu_1\nu_2^\ell\nu_3$. The
parity of the lower level is even ($e$) and odd ($f$). The
transitions are labelled as R, P, and Q for
$J_{up}-J_{low}=+1$, $-1$, 0.}
\label{tab:frequencies_sis}
\end{deluxetable}

In order to qualitatively interpret the effects of IR overlaps on maser
emission, we have used the same LVG radiative transfer code
modified to account for overlaps,
changing the intensity at the overlapping frequency and
the escape probability for photons from the overlapped lines.
Thus, for some SiS lines, the excitation temperature becomes negative
(maser activation)
and the brightness temperature is considerably enhanced.
M$_2$ and M$_3$ are naturally explained by the
overlap of the R$\left(13\right)$ line of SiS with the
strong C$_2$H$_2$ line $0^01^1-0^00^0$R$_e\left(9\right)$
(Table~\ref{tab:frequencies_sis}). SiS molecules are easily excited
from $v$=0 $J$=13 to $v$=1 $J$=14 through R$\left(13\right)$ and
decay to the $v$=0 $J$=15 via P$\left(15\right)$,
creating a population inversion between the $v$=0 $J$=13 and
15, and producing maser emission in M$_2$ and M$_3$.
M$_1$ may be produced by the overlap of
the P$\left(10\right)$ SiS line with the strong C$_2$H$_2$ line
$1^{-1}1^1-1^{-1}0^0$R$_f\left(3\right)$. This overlap
can pump SiS molecules from $v$=0 $J$=10 to $v$=1 $J$=9
through P$\left(10\right)$,
depopulates the $v$=0 $J$=10 level and produces the inversion
between $v$=0 $J$=10 and 11.

We have also looked for overlaps of $v$=1-0
higher-$J$ SiS lines with C$_2$H$_2$ and HCN transitions
to try to predict SiS masers at submillimeter
wavelengths.
Some of them are shown in the second block of
Table~\ref{tab:frequencies_sis}. They suggest, for example, that a
maser could be found in rotational transitions involving the $v$=0
$J$=23 level (likely $J$=24-23 and maybe 23-22), or the $v$=0
$J$=25 and 26 states (perhaps in the $v$=0 $J$=27-26 and maybe
26-25). These overlaps could also produce masers in $v$=1
rotational transitions.

There are many overlaps of other SiS isotopologues with
lines of C$_2$H$_2$, HCN and $^{28}$Si$^{32}$S.
With the adopted criteria for the overlap search (see footnote
of Table~\ref{tab:frequencies_sis}),
we have found 91, 93, 76, and 94 coincidences for $^{29}$SiS,
$^{30}$SiS, Si$^{34}$S, and Si$^{33}$S, respectively. Consequently,
although these species are less abundant than
$^{28}$Si$^{32}$S, the population of some levels could
be inverted producing maser emission.

This study represents the discovery of three new SiS
masers and should be complemented with future observations
of higher-$J$ $v$=0 and 1 rotational transitions.
Furthermore, a detailed multi-molecule
non-local radiative transfer model would help to understand
the dust formation region and the role of SiS in C-rich
evolved stars.

\acknowledgments

We want to thank the IRAM staff, M. J. Richter (NSF grant
AST-0307497), J. H. Lacy (NSF grant AST-0205518), and collaborators for
the mid-infrared observations,
and spanish MEC for funding support
through grant ESP2004-665, AYA2003-2785, and ``Comunidad de Madrid''
Government under
PRICIT project S-0505/ESP-0237 (ASTROCAM). This study is
supported in part by the European Community's human potential
Programme under contract MCRTN-CT-2004-51230, Molecular Universe.

\end{document}